\documentclass[fleqn,usenatbib]{mnras}
\usepackage[T1]{fontenc}
\usepackage{ae,aecompl}
\usepackage{multirow}
\usepackage{xcolor}
\usepackage{mathtools}
\usepackage{graphicx}	
\usepackage{amsmath}	
\usepackage{amssymb}	
\newcommand{\bz}{$\langle B_z \rangle$}

\usepackage{newtxtext,newtxmath}

\title[Upper cut-off frequency of ECME from MRPs]{What leads to premature upper cut-off frequencies of auroral radio emission from hot magnetic stars?}

\author[B. Das et al.]{
Barnali Das$^{1}$\thanks{E-mail: barnali@udel.edu},
Poonam Chandra$^{2,3}$
and
V\'eronique Petit$^{1}$
\\\\
$^{1}$Department of Physics and Astronomy, Bartol Research Institute, University of Delaware, 217 Sharp Lab, Newark, DE 19716, USA\\
$^{2}$National Centre for Radio Astrophysics, Tata Institute of Fundamental Research,  Pune University Campus, Pune-411007, India\\
$^{3}$National Radio Astronomy Observatory, 520 Edgemont Rd, Charlottesville, VA 22903, USA\\
}

\begin{document}
\label{firstpage}
\pagerange{\pageref{firstpage}--\pageref{lastpage}}
\maketitle

\begin{abstract}
Recently a large number of hot magnetic stars have been discovered to produce auroral radio emission by the process of electron cyclotron maser emission (ECME). Such stars have been given the name of Main-sequence Radio Pulse emitters (MRPs). The phenomenon characterizing MRPs is very similar to that exhibited by planets like the Jupiter. However, one important aspect in which the MRPs differ from aurorae exhibited by planets is the upper cut-off frequency of the ECME spectrum. While Jupiter's upper cut-off frequency was found to correspond to its maximum surface magnetic field strength, the same for MRPs are always found to be much smaller than the frequencies corresponding to their maximum surface magnetic field strength. 
In this paper, we report the wideband observations (0.4--4.0 GHz) of the MRPs HD\,35298 that enabled us to locate the upper cut-off frequency of its ECME spectrum. This makes HD\,35298 the sixth MRP with a known constraint on the upper cut-off frequency. With these information, for the first time we investigate into what could lead to the premature cut-off. We review the existing scenarios attempting to explain this effect, and arrive at the conclusion that none of them can satisfactorily explain all the observations. We speculate that more than one physical processes might be in play to produce the observed characteristics of ECME cut-off for hot magnetic stars. Further observations, both for discovering more hot magnetic stars producing ECME, and to precisely locate the upper cut-off, will be critical to solve this problem.
\end{abstract}

\section{Introduction}\label{sec:intro}
A wide varieties of magnetic celestial objects emit auroral radio emission, starting from the cool planets to the very hot upper main-sequence stars. The emission mechanism behind is the electron cyclotron maser emission (ECME). ECME is intrinsically a narrow bandwidth phenomenon with the frequency of emission being proportional to the local magnetic field strength. That is why the emission is considered a highly reliable estimator of magnetic field in the emitting body. The observed ECME radiation, however, often consists of emission over a wide range of frequencies. This happens because in a magnetic star, different regions have different magnetic field strengths, and accordingly the different sites produce ECME at different frequencies. The regions closer to the star (where the magnetic field strength is usually higher) produce ECME at higher frequencies than that produced by regions farther from the star. This collectively gives rise to a broadband emission. The lower cut-off frequency of the observed spectrum may arise due to unavailability of sufficient number of non-thermal electrons (relative to the number of thermal electrons) at far-away location from the emitting object (where the magnetic field is weaker). On the other hand, the upper cut-off is expected to be governed by the maximum surface magnetic field of the emitting body. This has been found to be the case for Jupiter \citep[e.g.][]{zarka2004}, and often used as an indirect method to estimate the polar magnetic field strengths in ultracool dwarfs \citep[UCDs, e.g.][]{hallinan2006}. However, there is one class of objects that clearly violate this scenario. These are the magnetic massive stars producing ECME, also known as the `Main-sequence Radio Pulse emitters' \citep[MRPs,][]{das2021a}. In case of MRPs, the magnetic fields are already well-measured through spectropolarimetric observations \citep[e.g.,][]{shultz2018,shultz2019b,shultz2019c}. In most cases, the magnetic data are consistent with the stars having a magnetic field with near-dipolar topology with polar strengths ranging between $10^2-10^5$ gauss \citep[e.g.][etc.]{petit2013,shultz2018}. So far, fifteen MRPs have been discovered \citep{trigilio2000,chandra2015,das2018,lenc2018,leto2019,das2019a,das2019b,leto2020,leto2020b,das2022}, and it has been speculated that majority of the hot magnetic stars are probably MRPs \citep{das2022}. Among them, eleven were first discovered to produce ECME at sub-GHz frequencies \citep{chandra2015,das2018,lenc2018,das2019a,das2019b,das2022}. In fact, the highest frequency at which ECME has been confirmed from an MRP is $\lesssim 8$ GHz \citep[from the MRP CU\,Vir,][]{das2021a}, which corresponds to a field strength of 2.8 kG for emission at the fundamental harmonic. Among the remaining MRPs, the upper cut-off frequencies of ECME are constrained for only four of them (HD 133880, HD 142301, HD 147933 and HD 147932). In all cases (including CU\,Vir), the ratio of the upper cut-off frequency to the electron gyrofrequency corresponding to the maximum observed magnetic field strength is significantly smaller than unity (Table \ref{tab:cut_off}). Coincidentally, the lowest value of the constraint on the ECME upper cut-off is obtained for the MRP with the highest surface magnetic field strength (HD\,142301, Table \ref{tab:cut_off}).
Although a firm conclusion can only be drawn with a precise localisation of the cut-off frequencies, it is now clear that the reason that leads to the cut-off in the upper end of the ECME spectrum is different from that of the planets. What leads to the premature cut-off at the higher end of the observed ECME spectra for MRPs, remains unsettled.

\begin{table*}
\tiny
    \centering
    \caption{Table comparing the available constraints on the upper cut-off frequency ($\nu_\mathrm{upper}$) of ECME from MRPs with their maximum surface magnetic field strength $B_\mathrm{max}$ (obtained from spectropolarimetry), and the corresponding frequency $\nu_\mathrm{max}$($=2.8 B_\mathrm{max}$). Also shown are the upper limits to the magnetic field strength ($B_\mathrm{upper}$) at the emission site corresponding to the upper cut-off frequency, (assuming emission at the fundamental harmonic), and the respective estimates of the radial distances $r_\mathrm{min}$ (assuming a dipolar magnetic field). The columns, labelled `Inclination angle' and `Obliquity' show the angles made by the stellar rotational axis with the line-of-sight and the magnetic dipole axis respectively.
    }
    \begin{tabular}{ccccccccc|l}
    \hline\hline
         Star & $\nu_\mathrm{upper}$ & $B_\mathrm{max}$ & $\nu_\mathrm{max}$ & $\nu_\mathrm{upper}/\nu_\mathrm{max}$ & $B_\mathrm{upper}$ &$r_\mathrm{min}$ & Inclination & Obliquity & Reference\\
         & (GHz) & (kG) & (GHz) &($f$) & (kG) & ($R_*$) & angle & &\\
         \hline 
         CU\,Vir & $5<\nu_\mathrm{upper}\lesssim 8$ & $3.8\pm 0.2$ & $10.6\pm 0.6$ & $0.45<f\lesssim 0.8$ & 2.8 & 1.1& $46.5^\circ \pm 4.1^\circ$ &$79^\circ\pm 2^\circ$  &\citet{leto2006,das2021a} \\
         & & & & & & & & &\citet{kochukhov2014}\\
         HD\,133880 & $4<\nu_\mathrm{upper}< 5$ & $ 9.6\pm1 $ & $26.9\pm 2.8$ & $0.13<f<0.21$ & 1.8 & $1.7$ &$55^\circ\pm 10^\circ$ & $78^\circ\pm 10^\circ$ &\citet{lim1996,bailey2012}\\
         & & & & & & & & &\citet{das2020b,kochukhov2017}\\
         HD\,142301 & $1.5<\nu_\mathrm{upper}<5.5$ & $12.5^{+9}_{-0.3}$ & $35^{+25.2}_{-0.8}$ & $0.02<f<0.16$ & 2.0 & $1.8$ & $68^\circ\pm 5^\circ$ & ${58^\circ}^{+4^\circ}_{-12^\circ}$ &\citet{leto2019,shultz2020}\\\\
         HD\,147933 & $2.1<\nu_\mathrm{upper}<5.5$ & $2.7^{+0.9}_{-0.7}$ & $7.6^{+2.5}_{-0.8}$ & $0.21<f<0.98$ & 2.0 & 1.1& ${35^\circ}^{+8}_{-6}$ & ${78^\circ}^{+5}_{-8}$ &\citet{leto2020}\\\\
         HD\,147932 & $2.1<\nu_\mathrm{upper}< 9$ & $10.4^{+13.1}_{-0.8}$ & $29.2^{+36.68}_{-2.2}$  & $0.03<f< 0.20$ & 3.2 & 1.5& ${64^\circ}^{+6^\circ}_{-4^\circ}$ & $7^\circ\pm2^\circ$ &\citet{alecian2014,leto2020b,rebull2018}\\
         & & & & & & & & &\citet{shultz2022}, Shultz et al. (in prep.)\\
         HD\,35298 & $\approx 5.2$ & $11.2\pm 1$ & $31.4\pm 2.8$ & $\approx 0.16\pm 0.02 $ & 1.8 & 1.8 & $64^\circ\pm 4^\circ$& $78^\circ\pm 2$ &\citet{shultz2019b,shultz2019c}, This work\\
         \hline 
    \end{tabular}
    \label{tab:cut_off}
\end{table*}

In this paper, we report the wideband observations of the MRP HD\,35298 that enables us to locate the upper cut-off frequency of ECME. Adding it to the five other MRPs listed on Table \ref{tab:cut_off}, we discuss the validity of the existing hypotheses that attempt to explain the occurrence of ECME upper cut-off in magnetic hot stars.  

Throughout the paper, we have used the IAU/IEEE convention for right and left circular polarization (RCP and LCP respectively).

This paper is structured as follows: in \S\ref{sec:observation}, we describe our wideband observations, and the data analysis process; this is followed by our results in \S\ref{sec:results}. These results are discussed in \S\ref{sec:discussion}, and we then present our main conclusions in \S\ref{sec:conclusion}.


\begin{table*}
{\scriptsize
\caption{Log of observations of HD\,35298 showing the dates and durations of observations at different wavebands and the effective observing frequency ranges (Eff. band used for our analysis after removing the edges of the band, and excluding the corrupted spectral windows) for each band on different days. \label{tab:obs}}
\begin{tabular}{ccccc||ccccc}
\hline\hline
\multicolumn{5}{c}{uGMRT band 3}       & \multicolumn{5}{|c|}{VLA L+S}\\
Date & HJD range & Eff. band & Flux & Phase &  Date & HJD range & Eff. band & Flux & Phase\\
 &$-2.45\times 10^6$ & (MHz) & Calibrator & Calibrator & &$-2.45\times 10^6$ & (MHz) & Calibrator & Calibrator\\
\hline\hline
 2019--08--11 & $8706.65\pm 0.10$ & 334--461 & 3C48, 3C286 & J0607--085 & 2020--11--19 & $9172.93\pm 0.08$ & 1039.5--1679.5, & 3C147 & J0532+0732\\
 & & & & & & & 2051--3947 & & \\
 2019--09--17 & $8744.52\pm 0.12$ & 334--461 & 3C48, 3C147 & J0607--085 & 2020--11--22 & $9171.83\pm 0.08$ & 1039.5--1103.5, 1359.5--1679.5 & 3C147 & J0532+0732\\
 & & & & & & & 2051--3563 & & \\
\hline
\end{tabular}
}
\end{table*}

\section{Observation and data reduction}\label{sec:observation}
HD\,35298 was observed at four frequency bands using two radio telescopes: the upgraded Giant Metrewave Radio Telescope (uGMRT) and the Karl G. Jansky Very Large Array (VLA). The uGMRT was used to observe over the frequency ranges of 300--500 MHz (band 3) and 550--750 MHz (band 4), and the VLA was used to observe over 1--2 GHz ($L$ band) and 2--4 GHz ($S$ band). The $L$ and $S$ bands observations were conducted employing the subarray mode. The band 4 observations for the star were already reported by \citet{das2019b}. 
Here we present the band 3, and $L+S$ bands observations, and combine them with the already published band 4 results.

The details of the observations are given on Table \ref{tab:obs}. The uGMRT data have single spectral windows at both frequency bands, whereas each of the VLA L and S bands are divided into sixteen spectral windows. The data were analyzed using the `Common Astronomy Software Applications' \citep[\textsc{casa},][]{mcmullin2007} following the procedure described in \citet{das2020b}.

The data for HD\,35298 were phased over the known rotation period using the ephemeris of \citet{shultz2018}.

\section{Results}\label{sec:results}
\begin{figure}
    \centering
    \includegraphics[width=0.4\textwidth]{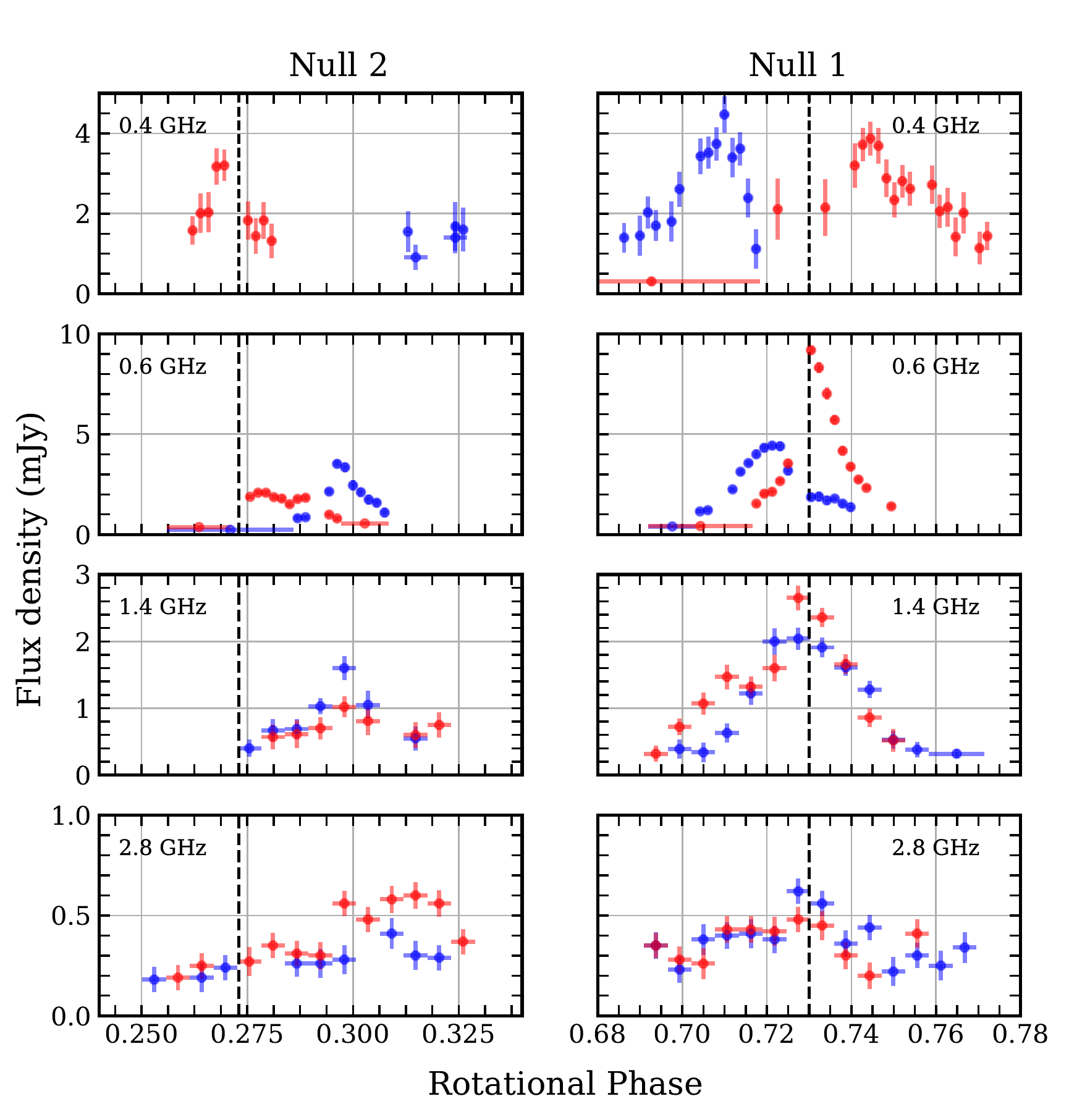}
    \caption{The lightcurves of HD\,35298 at different frequencies. The vertical dashed lines mark the magnetic nulls. The red and blue markers represent RCP and LCP respectively. Note that the data corresponding to the lightcurves at 0.6 GHz were already reported by \citet{das2019b}.}
    \label{fig:hd35298_lightcurves}
\end{figure}

\begin{figure*}
    \centering
    \includegraphics[width=0.85\textwidth]{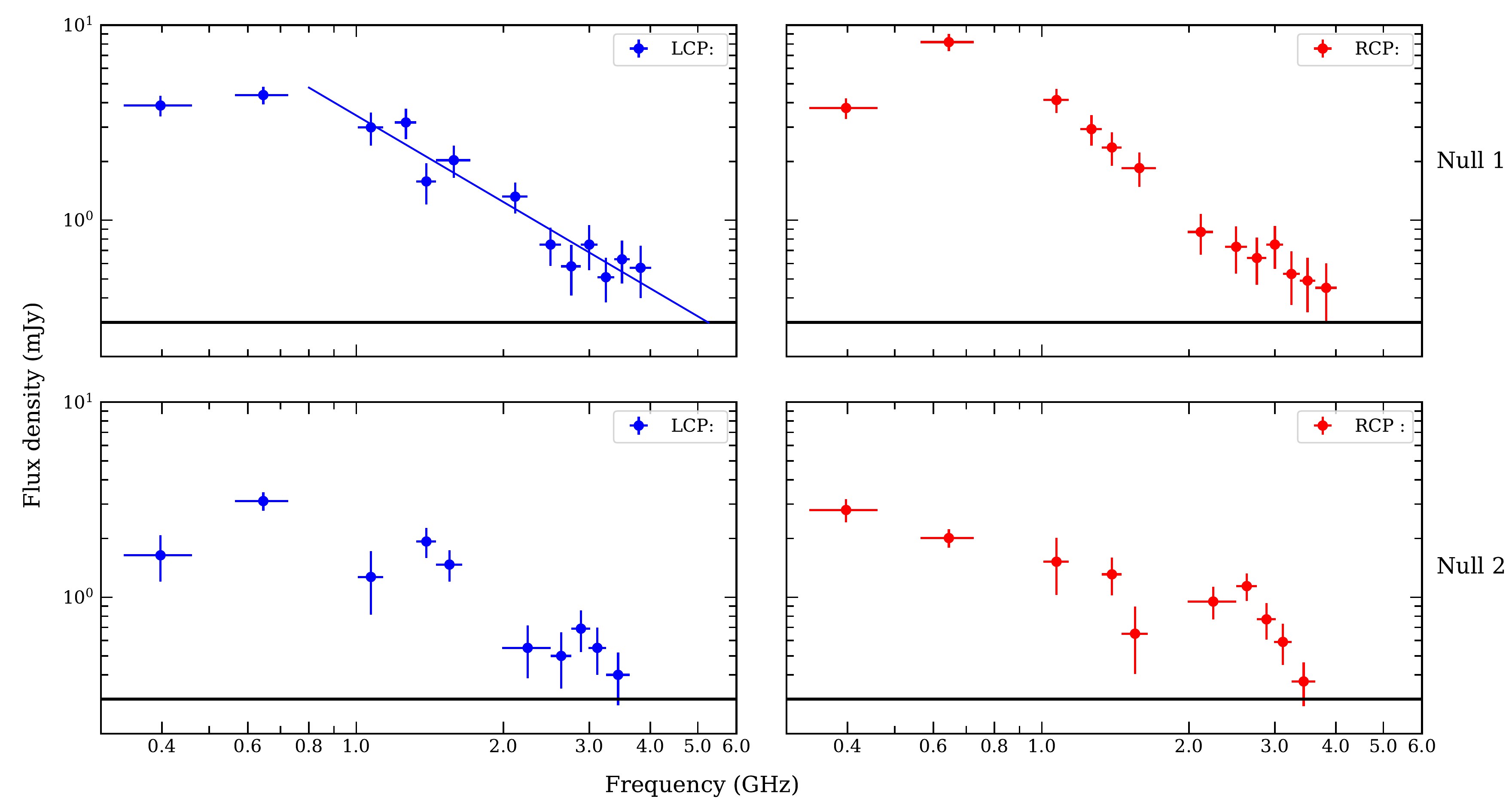}
    \caption{The peak flux density spectra of the ECME pulses observed for HD\,35298. The time resolution corresponding to each point is 15 minutes, except for the LCP data point at $\approx 0.4$ GHz near null at phase 0.27 (null 2), the time resolution in this case is 10 minutes. The horizontal line at each panel mark the approximate basal flux density at 1.4 GHz. For the LCP pulse near null 1 (top left panel), the upper cut-off frequency lies outside our observing frequency range. We estimated the upper cut-off frequency to be around 5.2 GHz by fitting a power law to the spectrum at and above 1 GHz (shown by the blue solid line).}
    \label{fig:hd35298_spectra}
\end{figure*}

The lightcurves of HD\,35298 near its two magnetic nulls are shown in Figure \ref{fig:hd35298_lightcurves}. As can be seen in Figure \ref{fig:hd35298_lightcurves}, the separation between the oppositely circularly polarized pulses decreases with increasing frequency such that above 1 GHz, the two pulses overlap. The sequence of arrival of the oppositely circularly polarized pulses at 0.4 GHz is the same as that observed at 0.6 GHz, and is consistent with X-mode emission \citep{das_erratum_hd35298}. 
The brightest pulses were observed at 0.6 GHz around the magnetic null at phase 0.7, which we now refer to as null 1 following the nomenclature introduced by \citet{das2019a}\footnote{Null 1 is the magnetic null phase where \bz~changes from negative to positive, and Null 2 is the magnetic null phase where \bz~changes from positive to negative. 
}. This is more evident from the peak flux density spectra shown in Figure \ref{fig:hd35298_spectra}. As in the case of HD\,133880 \citep{das2020b}, here also the spectra for the four pulses are not identical. Except for our highest frequency band (2--4 GHz), the ECME pulses observed near null 1 are brighter than the corresponding pulses observed near the other magnetic null. The spectra for the LCP pulses near null 2 (0.27 phase), and RCP pulses near null 1 exhibit clear turn-over over the frequency range of band 4 ($\approx 645$ MHz), and the spectrum for the LCP pulse near null 1 also exhibits signature of turn-over at around 0.6 GHz; however the spectrum for the RCP pulse near null 2 does not exhibit any sign of a turn-over down to our lowest frequency of observation ($\approx 0.4$ GHz). 

We now consider the upper cut-off frequency of ECME from HD\,35298. For that, we need an estimate of the basal flux density. Unfortunately, as the star is one of the farthest MRPs known, it has a very low basal flux density so that we could not obtain a lightcurve for the basal flux density variation as a function of frequency. We hence assign the minimum flux density observed at 1.4 GHz as an estimate of the basal flux density (see third right panel of Figure \ref{fig:hd35298_lightcurves}), which is $\approx 0.3$ mJy. We define the upper cut-off frequency as the lowest frequency at which, within the measurement uncertainty, the peak ECME flux density becomes consistent with the basal flux density. According to this definition, the upper cut-off frequency of both RCP and LCP pulses around null 2 is $\approx 3.4\,\mathrm{GHz}$, that for the RCP pulse near null 1 is $\approx 3.8$ GHz, but for the LCP pulse observable near 0.73 phase, the upper cut-off frequency lies outside our frequency range of observation. From the bottom left panel of Figure \ref{fig:hd35298_spectra}, we find that for this LCP pulse, the peak ECME flux density follows nearly a power-law above 1 GHz. To obtain an estimate of the frequency at which the peak flux density becomes 0.3 mJy (the basal flux density), we fitted a power-law to the data points above 1 GHz. This method returned a spectral index of $-1.5$, and the estimated upper cut-off frequency of 5.2 GHz.

\section{Discussion}\label{sec:discussion}
HD\,35298 is only the third MRP for which ultra-wideband observation of ECME are reported \citep[after HD\,133880 and CU\,Vir,][]{das2020a,das2021a}. Like the other two stars, it also has a large obliquity (angle between stellar rotation and magnetic dipole axes) of $\approx 78^\circ$ \citep{shultz2019c}. \citet{das2020b} proposed that such high obliquities may result into ECME properties being dependent on stellar orientation (e.g. different spectral properties for the pulses observed near the two magnetic nulls). 
This is due to the fact that the obliquity plays a key role in defining the distribution of stellar wind plasma in the magnetosphere \citep[the `Rigidly Rotating Magnetosphere', or RRM model,][]{townsend2005}. For aligned rotational and dipolar axes (obliquity$=0$), the plasma is accumulating over a disc at the magnetic equator, which is symmetric about the magnetic axis. However, as the obliquity increases, the plasma distribution losses its magnetic azimuthal symmetry. As a result of that, the ECME produced at different emission sites may encounter different plasma conditions along its path.
Since the radio pulses observed at different rotational phases (i.e. different stellar orientations) correspond to different sites of origin at the stellar magnetosphere, they are affected by the magnetospheric plasma differently, and to different extents,
which is manifested as different pulse properties at different rotational phases.
It is, however, to be kept in mind that the observations at band 3, band 4 and $L+S$ bands were obtained at different epochs. ECME pulses have been observed to exhibit variable pulse-height \citep[e.g.][]{trigilio2011}, and hence, in principle, difference between the spectra around the two nulls could also arise artificially due to the time-variable pulse properties.
This latter notion, however, contradicts the recent proposition by \citet{das2021a} where they suggested that ECME is an intrinsically stable phenomenon, and the variable pulse-height observed is a result of the centrifugal breakout events that lead to correlated change in the flux density of the pulses produced at the two magnetic hemispheres. 

The central result of this study is the estimation of the ECME upper cut-off frequency for HD\,35298: $\approx 5$ GHz, which has allowed us to investigate the existing explanations for ECME upper cut-off.
HD\,35298 has a maximum polar magnetic field strength of 10.8 kG \citep{shultz2019c}, corresponding to an electron gyrofrequency of 30.2 GHz. The ratio of the ECME upper cut-off frequency ($\approx 5$ GHz) to the maximum electron gyrofrequency is 0.17. Thus this MRP also suffers from a premature cut-off in the higher end of the ECME spectra like the other MRPs listed on Table \ref{tab:cut_off}. In the following subsection, we review the existing ideas regarding what could give rise to this effect.

\subsection{Existing scenarios attempting to explaining premature cut-off}\label{subsec:cut_off_idea1}
ECME frequency is proportional to the local electron gyrofrequency. As a result, higher frequency emission arises closer to the star (where the magnetic field strength is higher) and vice-versa. \citet{leto2019} proposed that the hot magnetic stars are likely to have high density thermal plasma close to their surface. As a result, below a certain height from the stellar surface, the plasma density is so high that the refractive index for the relevant frequency of emission becomes imaginary, which means the radiation cannot propagate, leading to a cut-off at a frequency that is lower than the frequency corresponding to the polar field strength. Alternately, high density plasma might inhibit the production of ECME itself by making the plasma frequency much larger than the electron gyrofrequency \citep[e.g.][]{melrose1982,treumann2006}.

We first consider the latter possibility where the premature \textit{cut-off is a result of the inhibition of the production of ECME by high density plasma at the auroral regions}. 
ECME is suppressed when the plasma frequency $\nu_\mathrm{p}\geq \nu_\mathrm{B}$, where $\nu_\mathrm{B}$ is the electron gyrofrequency. Let the radial coordinate of the region, in units of the stellar radius $R_*$, corresponding to the ECME upper cut-off frequency $\nu_\mathrm{upper}$ be $r_1$. Thus, in this scenario, for $r\leq r_1$, $\nu_\mathrm{p}/\nu_\mathrm{B}>=1$, with $\nu_\mathrm{p}(r_1)=\nu_\mathrm{B}(r_1)$. 

According to the scenario proposed by \citet{trigilio2004}, ECME is produced by non-thermal electrons in the middle magnetosphere, which is a thin transition region between the `inner' (where all the magnetic field lines are closed), and the `outer' (where all the magnetic field lines are open) magnetospheric regions. In this region, the magnetic field lines are stretched open near the magnetic equatorial region, but nevertheless, the field line topology close to the surface is nearly dipolar \citep[see Fig. 1 of ][]{trigilio2004}. 

Let us assume that the radial dependence of the magnetic field along the lines containing the ECME sites of production be $B\propto 1/r^b$, where $r$ is in units of the stellar radius ($R_*$), and $b$ lies between 2 and 3 ($b=3$ for pure dipolar field lines and $b=2$ for radial field lines).

If $s$ be the harmonic number of ECME, we have:
\begin{align}
   r_1&={\left[\frac{2.8s(B_\mathrm{max}/\mathrm{G})}{(\nu_\mathrm{upper}/\mathrm{MHz})}\right]}^\frac{1}{b}\label{eq:r1}
\end{align}
where $B_\mathrm{max}$ is the polar field strength (in units of gauss), and $\nu_\mathrm{upper}$ is in MHz. In order to find out how the plasma frequency varies with radial distance, we need to understand the radial dependence of the density $\rho$ in the auroral regions of the star. This can be obtained from mass continuity: $\rho A v_\mathrm{wind}=$ constant, where $A$ is the area of a magnetic flux tube and $v_\mathrm{wind}$ is the wind speed along that tube. From the magnetic flux conservation, we have $A\propto 1/B$, where $B$ is the local magnetic field. We have, $B\propto 1/r^b$, and $v_\mathrm{wind}$ is expressed as $v_\infty(1-1/r)^\beta$, $v_\infty$ is the wind terminal speed, and $\beta\sim 1$ \citep[e.g.][]{castor1975,owocki2004,trigilio2004}. Thus we have:

\begin{align}
    \rho v_\infty{\left(1-\frac{1}{r}\right)}^\beta & \propto B\propto \frac{1}{r^{b}}\nonumber\\
    \Rightarrow \rho & \propto \frac{1}{r^{b}{\left(1-\frac{1}{r}\right)}^\beta}\nonumber\\
    \Rightarrow \nu_\mathrm{p}\propto \sqrt{\rho}& \propto \frac{1}{\sqrt{r^{b}{\left(1-\frac{1}{r}\right)}^\beta}}\nonumber\\
   \Rightarrow \nu_\mathrm{p}&=\frac{\nu_\mathrm{p0}}{\sqrt{r^{b}{\left(1-\frac{1}{r}\right)}^\beta}} \label{eq:nu_p}
\end{align}

$\nu_\mathrm{p0}$ is a constant of proportionality which can be evaluated using the boundary condition $\nu_\mathrm{p}(r_1)=\nu_\mathrm{B}(r_1)$ as:
\begin{align*}
    \nu_\mathrm{p0}&=\nu_\mathrm{B}(r_1)\sqrt{r_1^{b}{\left(1-\frac{1}{r_1}\right)}^\beta}
\end{align*}
$r_1$ is given by Eq. \ref{eq:r1}, and $\nu_\mathrm{B}(r_1)=\nu_\mathrm{upper}/s$. Substituting these in Eq. \ref{eq:nu_p} gives:
\begin{align}
    \nu_\mathrm{p}&=\frac{\nu_\mathrm{upper}}{s}{\left(\frac{r_1}{r}\right)}^\frac{{b}}{2}{\left[\frac{1-(1/r_1)}{1-(1/r)}\right]}^\frac{\beta}{2} \label{eq:nu_p_r_variation}
\end{align}
Thus, Eq. \ref{eq:nu_p_r_variation} will give the radial dependence of the plasma frequency. The same for the electron gyrofrequency is given by $2.8B_\mathrm{max}/r^{b}$ (in MHz when $B_\mathrm{max}$ is in gauss units). We then have (and using the fact that $\nu_\mathrm{upper}=2.8sB_\mathrm{max}/r_1^{b}$):

\begin{align}
    \frac{\nu_\mathrm{p}}{\nu_\mathrm{B}}&=\frac{\nu_\mathrm{upper}r^{b}}{2.8sB_\mathrm{max}}{\left(\frac{r_1}{r}\right)}^\frac{{b}}{2}{\left[\frac{1-(1/r_1)}{1-(1/r)}\right]}^\frac{\beta}{2}\nonumber\\
    &={\left(\frac{r}{r_1}\right)}^\frac{{b}}{2}{\left[\frac{1-(1/r_1)}{1-(1/r)}\right]}^\frac{\beta}{2}\label{eq:nup_nuB}
\end{align}
Eq. \ref{eq:nup_nuB} gives the radial variation of the quantity $\nu_\mathrm{p}/\nu_\mathrm{B}$ that obeys the condition that at a radial distance $r=r_1$, the ratio equals unity. 
Under the scenario that the upper cut-off is due to the presence of high density plasma at the emission sites, we need to have $\nu_\mathrm{p}/\nu_\mathrm{B}>1$ for $r<r_1$ at the auroral regions.

\begin{figure*}
    \centering
    \includegraphics[width=0.33\textwidth]{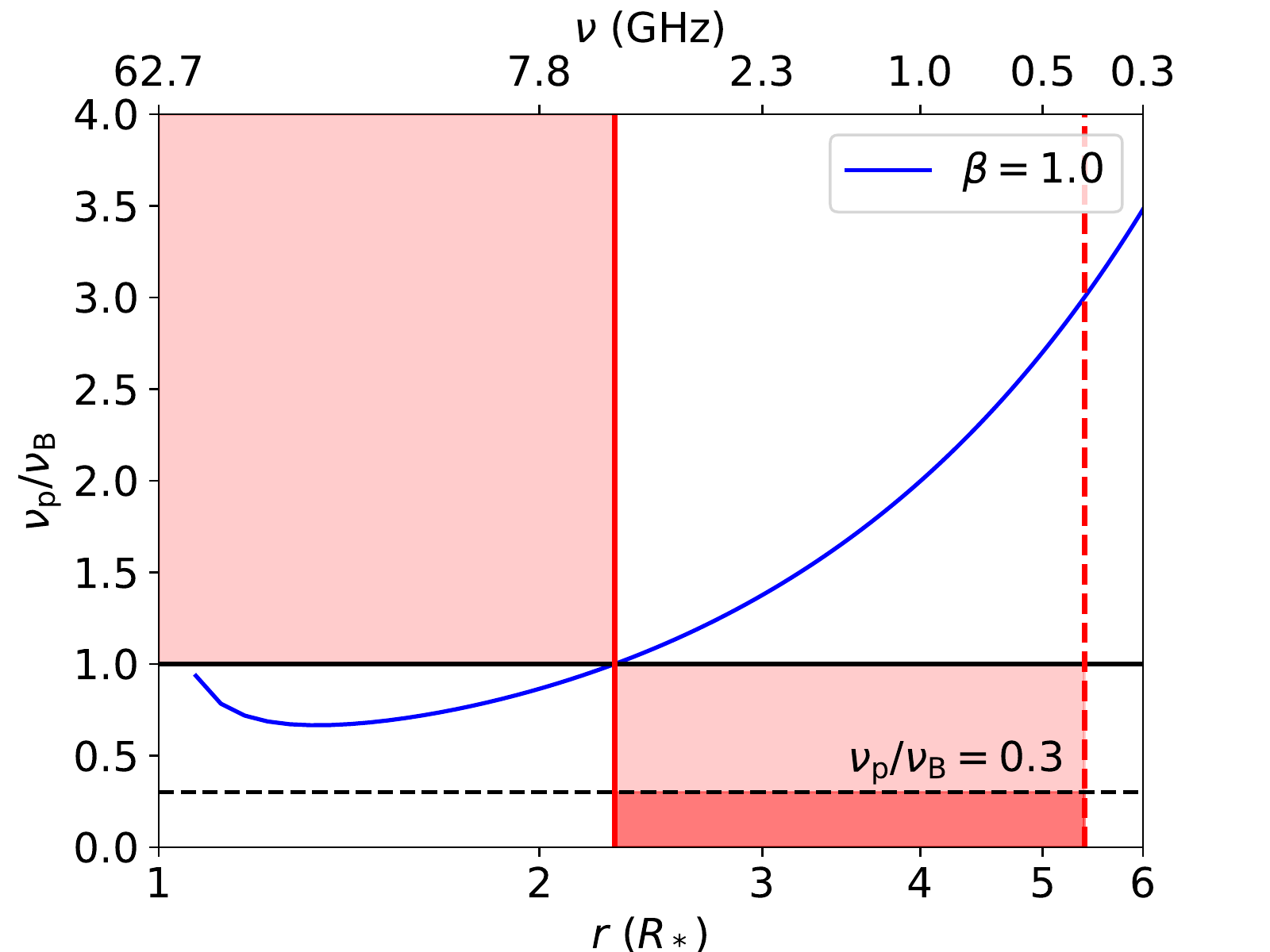}
    \includegraphics[width=0.33\textwidth]{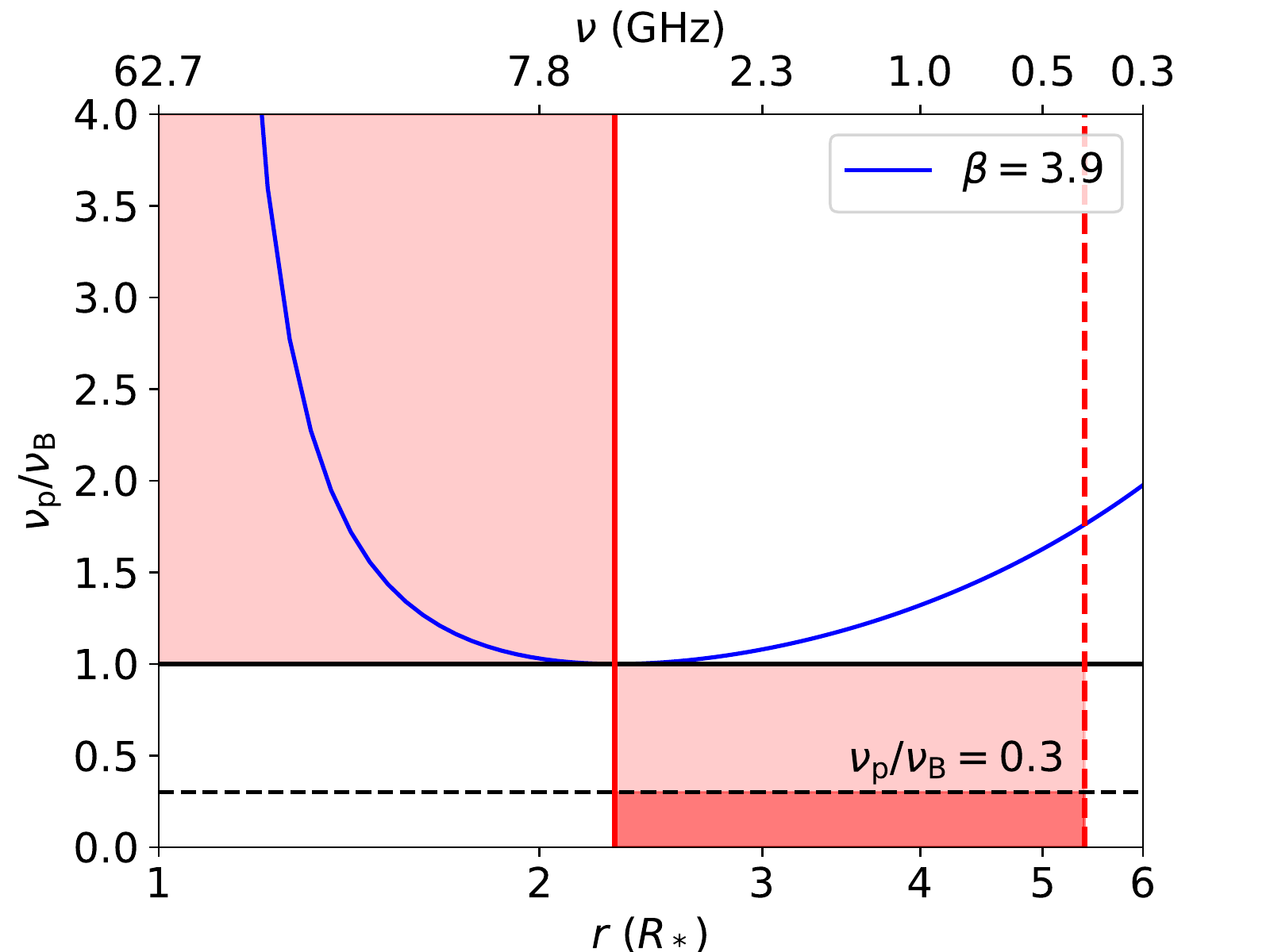}
    \includegraphics[width=0.33\textwidth]{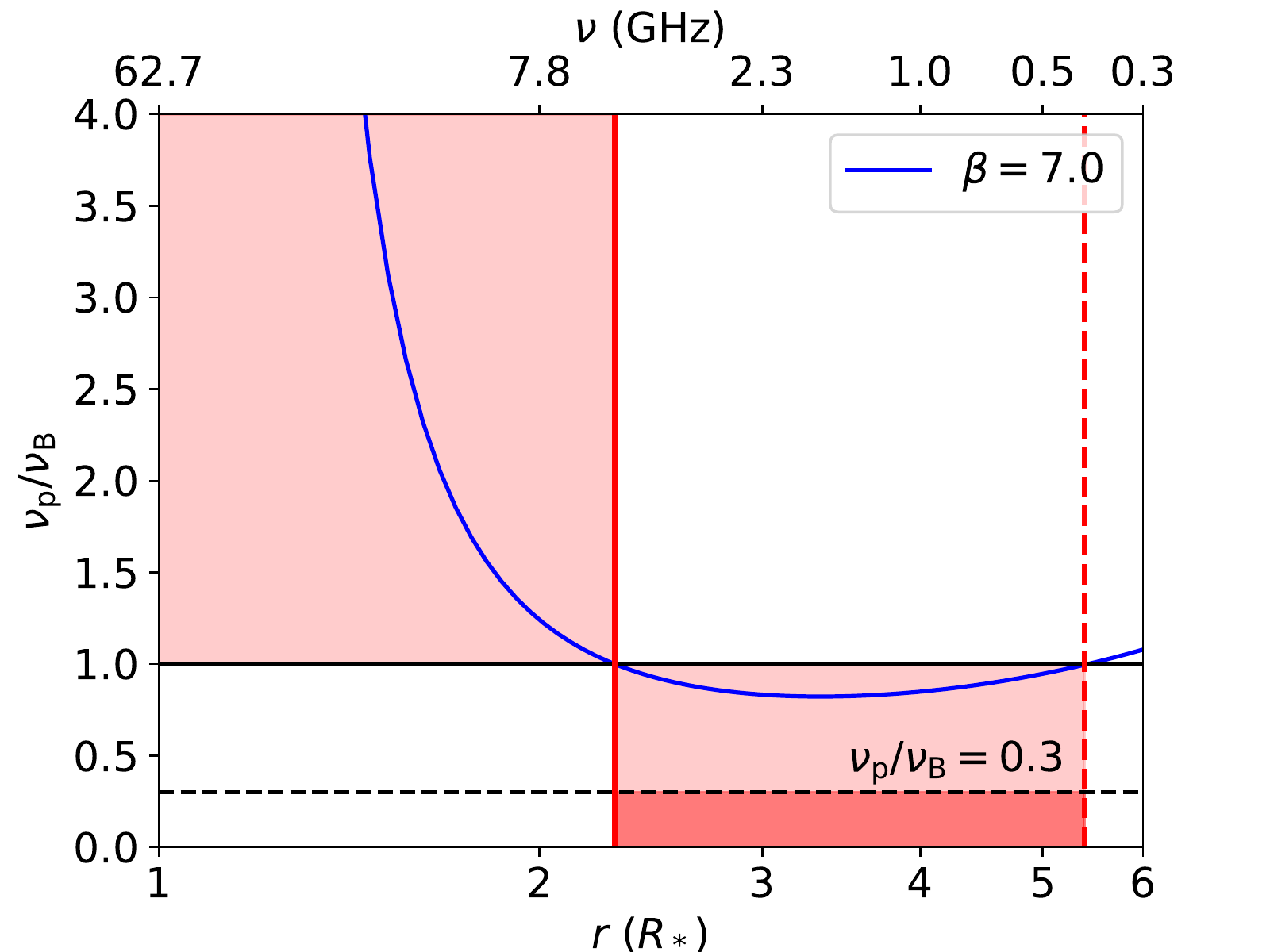}    
    \caption{The radial variation of the ratio of the plasma frequency $\nu_\mathrm{p}$ to the electron gyrofrequency $\nu_\mathrm{B}$ for three different values of the exponent $\beta$ in the velocity law (see \S\ref{subsec:cut_off_idea1}), for the star HD\,35298 assuming that at the height where the ECME cut-off occurs ($r=2.29\,R_*$, shown by the vertical solid red line in each panel), $\nu_\mathrm{p}=\nu_\mathrm{B}$. The harmonic number for ECME is taken to be $s=2$. The corresponding ECME frequencies are shown in the top X-axis. The red, dashed vertical line corresponds to the lowest observed frequency of ECME (0.4 GHz). The dashed horizontal line marks the approximate value of the ratio below which the magneto-ionic mode of emission is extra-ordinary \citep[e.g.][]{sharma1984,leto2019}. We expect the blue curve to lie in the shaded region for observation of ECME at 0.4--5.2 GHz which is satisfied only for $\beta\geq 7.0 $ (rightmost panel). A much higher value of $\beta$ will be needed if we further impose the condition that the magneto-ionic mode of emission should be extra-ordinary (the darker shaded regions in the panels).}
    \label{fig:hd35298_nup_nub}
\end{figure*}

We now consider the star HD\,35298, for which $B_\mathrm{max}=11200$ G and $\nu_\mathrm{upper}=5200$ MHz. 
Using these values and setting $b=3,\,s=2$ in Eq. \ref{eq:r1}, we get $r_1=2.29\,R_*$.
The left panel of Figure \ref{fig:hd35298_nup_nub} shows the radial variation of $\nu_\mathrm{p}/\nu_\mathrm{B}$ for $\beta=1$. Clearly, the resulting variation is inconsistent with the scenario ($\nu_\mathrm{p}/\nu_\mathrm{B}>=1$ for $r\leq r_1$ marked by the solid red vertical line) for $\beta\sim 1$. It can be shown that, $\beta$ needs to be $>{b}(r_1-1)$ to satisfy the scenario, which translates to $\beta>3.9$ (for $b=3$) for $r_1=2.29\,R_*$ (middle panel of Figure \ref{fig:hd35298_nup_nub}). Had we taken $s=1$, we would get $r_1=1.8$ and hence the necessary condition would be $\beta>2.4$. 
For $b=2$, the corresponding lower limit on $\beta$ are 4.9 and 2.9 for emission at the second and first harmonic respectively.

A stronger constraint on $\beta$ is obtained by incorporating the fact that the star emits ECME down to 400 MHz. If the corresponding radial distance is $r_0$ ($>r_1$), and we impose the condition that $\nu_\mathrm{p}/\nu_\mathrm{B}\leq 1$ at $r=r_0$, we get from Eq. \ref{eq:nup_nuB}:
\begin{align*}
    \beta& \geq {b}\frac{\log(r_0/r_1)}{\log{\left[\frac{1-(1/r_0)}{1-(1/r_1)}\right]}} 
\end{align*}
Using $b=3$ again, we find the above equation to translate
to $\beta\geq 7.0$ for emission at the second harmonic ($s=2$, right panel of Figure \ref{fig:hd35298_nup_nub}), and $\beta\geq 4.8$ for emission at the fundamental ($s=1$). 
The corresponding values of the lower limit to $\beta$ become even higher for $b=2$.
To the best of our knowledge, such high value of $\beta$ has not been proposed yet. In addition, if we now consider the fact that the observed magneto-ionic mode of ECME from HD\,35298 corresponds to extra-ordinary, we will need to consider an even higher value of $\beta$ to make $\nu_\mathrm{p}/\nu_\mathrm{B}<0.3$ \citep[e.g. ][]{sharma1984,leto2019}. These strong requirements disfavour this scenario.

We next consider the hypothesis that \textit{the cut-off is a result of the prohibition of the propagation of the radiation due to encountering high density plasma on its way through the stellar magnetosphere}. Such high density plasma can be present in the inner magnetospheres (the part of the stellar magnetosphere where the magnetic field energy is higher than the wind kinetic energy, and the magnetic field lines are closed) of stars with centrifugal magnetospheres (CMs). CMs refer to stellar magnetospheres in which the extent of the largest closed magnetic field line (represented by the Alfv\'en radius $R_\mathrm{A}$) is larger than the distance at which the centrifugal force due to co-rotation balances gravity \citep[called the Kepler radius $R_\mathrm{K}$;][]{petit2013}.
According to the RRM model \citep{townsend2005}, in those cases, stellar wind plasma can accumulate in the region between $R_\mathrm{K}$ and $R_\mathrm{A}$ leading to a region with very high plasma density. As mentioned already at the beginning of this section, the obliquity plays an important role in determining the plasma distribution in the CM.
When the magnetic and rotation axes are aligned (obliquity $=0$), the high density plasma accumulates at the magneto-rotational equator, forming a thin disc-like structure surrounding the star. This distribution becomes significantly complex and non-intuitive as the obliquity increases \citep[and need not exhibit symmetry about the magnetic axis, which might cause the cut-off frequencies of ECME to be a function of stellar orientation,][]{townsend2005,das2020a,das2020b}. 
In our sample of MRPs, HD\,147932 is the only star with nearly zero obliquity \citep[][Shultz et al. in prep., see Table \ref{tab:cut_off}]{leto2020b}. For the rest of the stars, the obliquities are large (Table \ref{tab:cut_off}). This makes it non-trivial to predict
whether or not the radiation will encounter the high density region, as that will require a precise determination of the stellar plasma distribution (considering the individual stellar physical parameters). For the case of HD\,147932, the ECME upper cut-off frequency is $<9$ GHz \citep{leto2020b}. Thus, the minimum height of ECME production (assuming emission at the fundamental harmonic) is $>$0.6 $R_*$ (from the stellar surface).
This implies that, if the high density region indeed lies at the magnetic equator, it must have a thickness larger than the stellar radius. Now, from the RRM framework, we can estimate the scale-height of the disc
by considering that the wind materials are distributed along the field lines due to hydrostatic stratification with near isothermal condition.
At the Kepler radius $R_\mathrm{K}$ (in units of the stellar radius), the scale-height is given by \citep{townsend2005}:
\begin{align*}
    h&=R_\mathrm{K}\sqrt{\frac{2kTR_*}{\mu G M_*}R_\mathrm{K}}\\
    &=R_\mathrm{K}\sqrt{\frac{2kT}{(G M_*\mu)/(R_*R_\mathrm{K})}}
\end{align*}
where $T$ is the temperature of the accumulated plasma, $M_*$ and $R_*$ are the stellar mass and radius, $\mu$ is the mean molecular weight and $G$ is the universal gravitational constant.
Thus the scale-height in units of the dimensionless Kepler radius ($R_\mathrm{K}$) is a function of the ratio between the thermal energy and the gravitational binding energy at the Kepler radius \citep{townsend2005}. In the photospheres of early-type stars, the quantity $kTR_*/\mu G M_*$, denoted by $\epsilon_*$, is $\approx 0.001$ \citep{townsend2005},
$R_\mathrm{K}\approx 1.94$ for HD\,147932 \citep[by taking $M_*=4.8\,M_\odot$, $R_*=3.3\,R_\odot$ and rotation period $P_\mathrm{rot}=0.8639$ days,][]{shultz2022,leto2020b,rebull2018}.

This gives $h=0.06R_\mathrm{K}=0.12 R_*$. Note that as $r$ increases beyond $R_\mathrm{K}$, the scale-height at the magnetic equator decreases \citep{townsend2005}.
Thus, the width of this disc, if it indeed remains at the magnetic equator, is inadequate to give rise to the cut-off. Nevertheless, we cannot rule out the scenario with the currently available information \citep[since the value of the obliquity is not exactly zero,][Shultz et al. in prep]{leto2020b}.


\subsection{Need for a new scenario}\label{subsec:new_idea}
The preceding subsection shows that the existing scenarios that attempt to explain the occurrence of the premature cut-off of ECME from MRPs are not satisfactory.
The available data however, already exhibits a trend totally unexpected from either scenario.
Table \ref{tab:cut_off} shows that within the uncertainties, the upper cut-off frequencies of all the MRPs are nearly identical. 
Neither of the above scenarios (discussed in \S\ref{subsec:cut_off_idea1}) provide an explanation for this observation. 
This could, however, be simply a limitation of the current sample of MRPs with known constraints on the upper cut-off frequencies produced by them. As can be seen from Table \ref{tab:cut_off}, the magnetic field range varies only within a factor $\lesssim 5$. Considering the uncertainties associated with the magnetic field measurements (which are sometimes the lower limits to the real uncertainties since they were obtained for a fixed value of the inclination angle), and those associated with the upper cut-off frequency estimates, our inference is only suggestive, and we are far from being able to make a definitive conclusion. Nevertheless, the unexpected possibility of the cut-off frequency being indifferent to the polar field strength further enhances the need to obtain more observation of MRPs (which span a larger range of magnetic field strength) over a wide frequency range.

In the case that the upper cut-off frequencies indeed turn out to be independent of the magnetic field strength, we will have to revisit our ideas about what causes the premature cut-off in the case of these hot magnetic stars.
One possible scenario that can explain this phenomenon could be the energy loss via gyrosynchrotron emission. It is known that the total power emitted by an electron gyrating in a magnetic field is proportional to the square of the magnetic field strength. Thus, at regions with high magnetic field strength, the non-thermal electrons lose their energy quickly so that there may not be sufficient energetic electrons at sites with high magnetic field strength to give rise to ECME at those high frequencies. A limitation of this scenario is that it cannot explain why the observed upper cut-off frequency is different for different ECME pulses. However, one thing to be kept in mind is that none of the hot magnetic stars under consideration has an axi-symmetric dipolar magnetic field aligned with the rotation axis, so that the sites of acceleration of electrons are not situated symmetrically at the magnetic equatorial regions. Alternatively, there could be more than one physical processes causing the observed `cut-off phenomenon': one that is behind the production of orientation dependent cut-off (such as the high density plasma in the inner magnetosphere), and the other that is behind switching off the ECME production completely above a certain frequency (e.g. excessive energy loss by non-thermal electrons at sites with high magnetic field).


\section{Conclusion}\label{sec:conclusion}
In this paper, we report the upper cut-off frequency of ECME from the MRP HD\,35298. With that, the total number of MRPs with known constraints on the ECME upper cut-off has become six. All six MRPs have cut-off frequencies that are smaller than the electron gyrofrequencies corresponding to their maximum surface magnetic field strengths. For the first time, we have attempted to test the current hypotheses put forward to explain this premature cut-off in hot magnetic stars. We reviewed the existing ideas and conclude that this effect is unlikely to be caused by the inhibition of the ECME production due to the presence of high density plasma at the auroral regions. Though we find that the other existing idea involving ECME radiation not being able to pass through the high density plasma in the inner magnetosphere, somewhat inconsistent with the notion of the density distribution in the inner magnetosphere, we are unable to confirm/rule out this alternate hypothesis with the currently available data. Finally,
the existing data seems to suggest that in the case of MRPs, the ECME upper cut-off frequencies are indifferent to the maximum surface magnetic field, a totally unexpected outcome lacking any plausible explanation at the moment.

One of the obvious obstacles towards understanding the phenomenon of premature cut-off is the small number of MRPs for which the upper cut-off is constrained. The number of MRPs itself is not large enough so as to consider it as a sample that reasonably spans the phase-space of physical properties of hot magnetic stars. For example, currently all the known MRPs can be considered as rapid rotators ($P_\mathrm{rot}<4\,\mathrm{days}$), making it impossible to investigate the role of rotation in the phenomenon of ECME. Thus, both wideband observations of known MRPs, and search for new MRPs will be crucial to understand what causes the premature upper cut-off in these hot magnetic stars.

\section*{Acknowledgements}
We acknowledge support of the Department of Atomic Energy, Government of India, under project no. 12-R\&D-TFR-5.02-0700. BD acknowledges support from the Bartol Research Institute. BD thanks Ana Gabela for giving her time to carefully inspect the manuscript for grammatical errors. BD thanks Surajit Mondal for useful discussions. VP acknowledges support by the National Science Foundation under Grant No. AST--1747658.
We thank the staff of the GMRT and the National Radio Astronomy Observatory (NRAO) that made our observations
possible. The GMRT is run by the National Centre for Radio Astrophysics of the Tata Institute of Fundamental Research. The National Radio Astronomy Observatory is a facility of the National Science Foundation operated under cooperative agreement by Associated Universities, Inc. This research has made use of NASA's Astrophysics Data System.

\section*{Data availability}
The uGMRT data used in this article are available in \url{https://naps.ncra.tifr.res.in/goa/data/search} under proposal codes 36\_034 and 34\_111. The VLA data are available in \url{https://archive.nrao.edu/archive/advquery.jsp} under the project code 20A-012. The data were analyzed using \textsc{casa} \citep{mcmullin2007}.


\bibliographystyle{mnras}
\bibliography{das} 




\bsp	

\label{lastpage}
\end{document}